# Resilient Cyber-Physical Systems: Using NFV Orchestration

**Jose Moura** [1*] **and David Hutchison** [2]

[1] School of Technology and Architecture, ISCTE-Instituto Universitário de Lisboa, 1649-026 Lisbon, Portugal, and Instituto de Telecomunicações, 1649-026 Lisbon, Portugal (e-mail: jose.moura@iscte-iul.pt)

[2] School of Computing and Communications, InfoLab21, Lancaster University, Lancaster LA1 4WA, U.K. (e-mail: d.hutchison@lancaster.ac.uk)

\* Correspondence: jose.moura@iscte-iul.pt

**Abstract:** Cyber-Physical Systems (CPSs) are increasingly important in critical areas of our society such as intelligent power grids, next generation mobile devices, and smart buildings. CPS operation has characteristics including considerable heterogeneity, variable dynamics, and high complexity. These systems have also scarce resources in order to satisfy their entire load demand, which can be divided into data processing and service execution. These new characteristics of CPSs need to be managed with novel strategies to ensure their resilient operation. Towards this goal, we propose an SDN-based solution enhanced by distributed Network Function Virtualization (NFV) modules located at the top-most level of our solution architecture. These NFV agents will take orchestrated management decisions among themselves to ensure a resilient CPS configuration against threats, and an optimum operation of the CPS. For this, we study and compare two distinct incentive mechanisms to enforce cooperation among NFVs. Thus, we aim to offer novel perspectives into the management of resilient CPSs, embedding IoT devices, modeled by Game Theory (GT), using the latest software and virtualization platforms.

## 1. Introduction

A Cyber-Physical System (CPS) is essentially a physical facility with embedded sensors and actuators that can be remotely monitored and controlled by computerized systems [1], which we assume here are distributed virtualized agents (e.g. NFVs), most of them located at the network edge. The monitoring and control of CPS are made by logical control loops over physical communication channels. These channels are established between the sensors/actuators and the NFVs. The channels transfer data representing the facility status and control messages to change the operation mode. CPSs are increasingly found in diverse applications areas such as power grids [2][3], smart buildings [4][5], next-generation mobile communication systems [6][7], healthcare systems [8][9], and precision farming systems [10][11].

Each NFV agent should take individual decisions based on some system contextual information. Nevertheless, each NFV agent has unique contextual information which may be different from what is available to others. Consequently, each NFV agent could make management decisions conflicting with the decisions from others. Thus, counter-balancing the flexibility of a distributed NFV approach, the CPS may have a sub-optimal performance when compared to centralized decision-making. This optimization inefficiency of the distributed management is like a system cost, representing a degradation of the CPS performance. To mitigate this performance degradation induced by the competition among NFV agents, we argue in favor of the utilization of a system mechanism to incentivize the cooperation and support orchestration amongst them.

This paper proposes and analyzes several incentives mechanisms for cooperation among virtualized agents to let them decide on a set of orchestrated management decisions that would enable resilient system operation in the presence of a serious threat or challenge. We also discuss reactive and proactive CPS orchestration mechanisms among the agents towards the fulfilment of global system goals. The paper has the following structure. Section 2 discusses related work. Section 3 presents a design of a software-defined resilient CPS. Section 4 evaluates our proposal. Section 5 concludes the paper and outlines future research. The paper's logical organization is visualized in Figure 1.

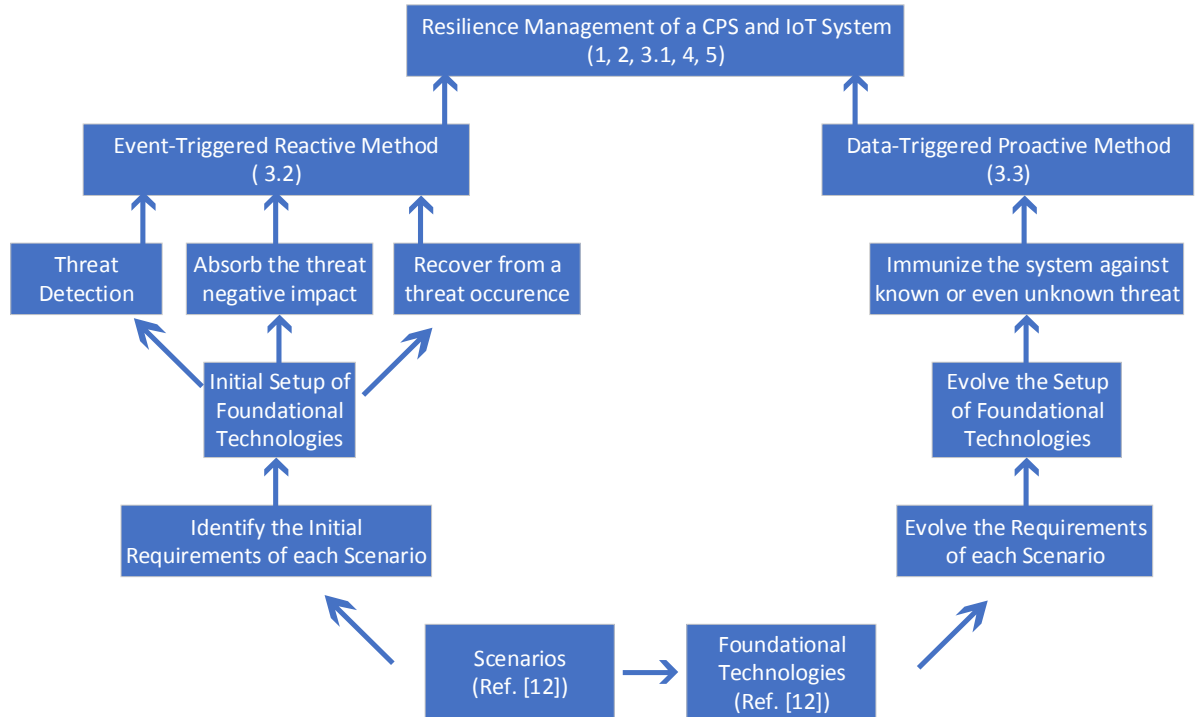

**Figure 1.** Logical Roadmap Behind the Paper

## 2. Related Work

The work in [12] presents a fundamental background in resilience and related concepts. It also offers a comprehensive discussion on diverse relevant scenarios for CPSs and on foundational technologies to enforce resilience in CPSs. Specifically, the authors of [13] discuss the state of the art in resilient networked systems.

Modern CPSs can be treated as large-scale heterogeneous distributed systems. In this context, when adequate supervision and control are also required for network-wide resilience, it is crucial to study the efficient orchestration [14] of a set of software-based services that must cooperate among themselves to fulfil the global resilience requirements [15]. Some software-based services that are pertinent to enforce wide-area resilient networked systems are pointed out in [15], such as traffic classification, anomaly detection, or traffic shaping.

The analysis of a resilient CPS can be made using a theoretical model. A very popular tool to perform system analysis is Game theory (GT) [16]. It is very useful for analyzing the diverse situations that could impair the system's normal operation. GT also enables the design of automatic models with either bounded rationality or decision uncertainty to safeguard the system's key functionalities in spite of the occurrence of serious threats [17]. In addition, the diverse model players should not only optimize their individual outcomes, but they should also coordinate among themselves towards the fulfilment of common global system goals. In our opinion, the efficient coordination among players can be globally guaranteed by correct incentives endorsed by the model players for them to cooperate among themselves. Aligned with these ideas, we have reviewed the literature for theoretical models in CPSs that incentivize cooperation among the players. This cooperation is fundamental to resilient CPSs, and it is discussed below.

The authors of [18] provide an in-depth literature review in sustainable incentives for mobile crowdsensing, discussing auctions, lotteries, and trust and reputation proposals. In contrast, [19] introduces a novel approach for mobile crowdsensing, viz. a social incentive mechanism, which incentivizes the coordinated positive contributions towards global system goals from the social friends of participants who perform sensing via their mobile devices.

In [18] the authors discuss the relevance of contract theory to design incentive mechanisms for use cases in wireless networks such as traffic offloading, spectrum trading, or mobile crowdsourcing.

We have found a considerable number of contributions addressing incentive models for cooperation among players in Vehicular Ad Hoc Networks (VANETs) to study the evolution of players' behavior (selfish vs. cooperative) under different network conditions [20][21], to motivate nodes to act as communication relays [22][23], and to influence nodes to support QoS-based communications [24]. A related survey is available in [25], which discusses several mechanisms to enforce cooperation – punishment based, misbehavior detection based, and mobile social networking based.

Further models to enforce cooperation within a system are as follows: i) hierarchical model [26][27][28]; ii) evolutionary model [29]; iii) cluster-based model [30]; and iv) potential game model [31][32]. In addition to these games, there is a mechanism design (or reverse GT) solution normally designated as auction model [33][34], which finds the optimum system status with a convergence time lower than that of a theoretical game [35]. Alternatively to the previous mechanisms which are based on (reverse) GT, [36] proposes an incentive mechanism based on both the anchoring effect and loss aversion of Behavioral Economics to stimulate data offloading in IoT use cases. The anchoring effect can be particularly useful, in the start of model game, when the players have not yet learned more suitable choices. In this way, the players are initially attracted to select a choice that optimizes the system operation (e.g. enforce nodes to perform data offloading across the existing edge computational resources, considering also the energy consumption / availability in each node).

The authors of [37] propose a virtual (based on NFV) and dynamic control (based on SDN) architecture. They deploy, at the SDN controller, a centralized non-cooperative incomplete information game. This enables the SDN controller to decide how the virtual (NFV) sensors are organized in clusters and to identify the more suitable sleep mode for each sensor. The final aim is to extend the lifetime of a software-defined CPS. Our current work is similar to [37] except the latter is concerned with energy efficiency and the former is towards the more efficient coordination among virtualized agents for supporting the CPS resilience in a more generic way. In addition, [38] is about a software-defined solution but without NFV. The authors of [39] survey the state of the art on the application of SDN and NFV to IoT use cases. In addition, [40] proposes a taxonomy of the evolution of the NFV/SDN relationship. Further, [41] reviews the literature on emerging SDN and NFV mechanisms for IoT Systems but mainly focused on the security aspects and not addressing the resilience feature.

The next section presents our basic design for a software-defined resilient CPS. It also discusses two more specific design options to orchestrate agent modules running over the SDN controller. The first option offers a reactive agent orchestration, and the second one a proactive agent orchestration.

## 3. Design of a Resilient Cyber-Physical System

This section discusses the design of a Cyber-Physical System (CPS) to enhance this with extra capabilities to detect, absorb and recover, and adapt against threats against the normal operation of each CPS [42]. We also discuss reactive and proactive orchestration mechanisms among the several CPS management entities towards the resilient operation of that CPS.

### 3.1. Design of a Software-Defined Resilient Cyber-Physical System

The current sub-section discusses some design aspects that are important to consider in a Software-Defined resilient CPS. Table I presents a four-layered hierarchical architecture [12], which can detect, absorb and recover, and adapt to threats made against CPSs [42]. Further details on this architecture are available in [12].

4 of 13

**Table I. Hierarchical Architecture of a Software-Defined Resilient Cyber-Physical System [12]**

| Layer | Plane | Domain | CPS Activity [42] | Goals | Tools |
|---|---|---|---|---|---|
| 4 | Intelligent management | Inter/Intra | Adapt | Reasoning, orchestration, full abstraction, adjust management policies or intents | NFV, SDN, GT, Intent Engine, ML/AI |
| 3 | Control | Intra | Adapt, recover | Partial abstraction, topology, traffic | Software-Defined Controller with link layer discovery, forwarding, and feedback loop |
| 2 | Switching | Edge | Detect, absorb, recover | Decision about next link decision, traffic mirroring, discard packet | OpenFlow rules in local device tables, queues |
| 1 | Physical communications | IoT | Detect, absorb | Accept or discard received message | Interface Chip programming |

The next sub-section debates the modeling of a software-defined resilient CPS, which is managed in a reactive way by each management agent.

*3.2. Design of a Reactive Software-Defined Resilient CPS Management Mechanism*

We discuss here the modeling of a solution to manage a resilient CPS in a reactive way. This solution has a four-layered design (see Table I). In addition, Figure 2 presents the key functional blocks of the system under investigation. Analyzing this, one can conclude that the CPS status is being supervised in a periodic way by the SDN controller via a Southbound API protocol such as OpenFlow (see Figure 2, message 1). Then, the SDN controller, acting as an intermediary, exchanges REST messages via Northbound API with the topmost level system agents (e.g. NFV modules). Using these messages, the SDN controller reports status events associated with the CPS operation. These events are analyzed, classified and processed. The agent decision depends on the processing of all the received events (see message 3, Figure 2). After, the agent decision is transferred to the SDN controller. Finally, the SDN controller converts the received management decision into flow rules that control the CPS physical infrastructure, also commonly referred as CPS data plane (see message 5, Figure 2).

The agent processing represented in Figure 2 as "Layer 4: Service Management", and shown in the topmost line of Table II, it is now briefly explained. Each individual agent estimates the system status from received event messages. The system status is evaluated as the ratio between the *Quantity_good_events* and the *Quantity_total_events*, both collected in a periodic way. There is also the estimator $S_n$, which is the system status at instant "n". This agent system status estimator with memory (i.e. configurable parameter $\alpha \in [0, 1]$) enables that agent to identify in the best way possible an eventual system anomaly and, after that, to react in a cooperative way to that issue. This means that the recover or adapt algorithm is executed by a specific agent if that agent decides to cooperate and if that choice has been randomly sorted out by the same agent – like tossing a coin. In addition, all the previous goals should be achieved by minimizing the usage of (heterogeneous) system resources (e.g. energy, bandwidth). Alternatively, the agent can selfishly select the 'defect' strategy. As the players select their strategies to optimize the system status, they then verify how the system behaves by evaluating the subsequent value for the local estimator of the system status, and the local processing in each agent is repeated as already explained. In parallel, the global system management is hopefully enhanced, increasing its robustness against any outcoming menace.

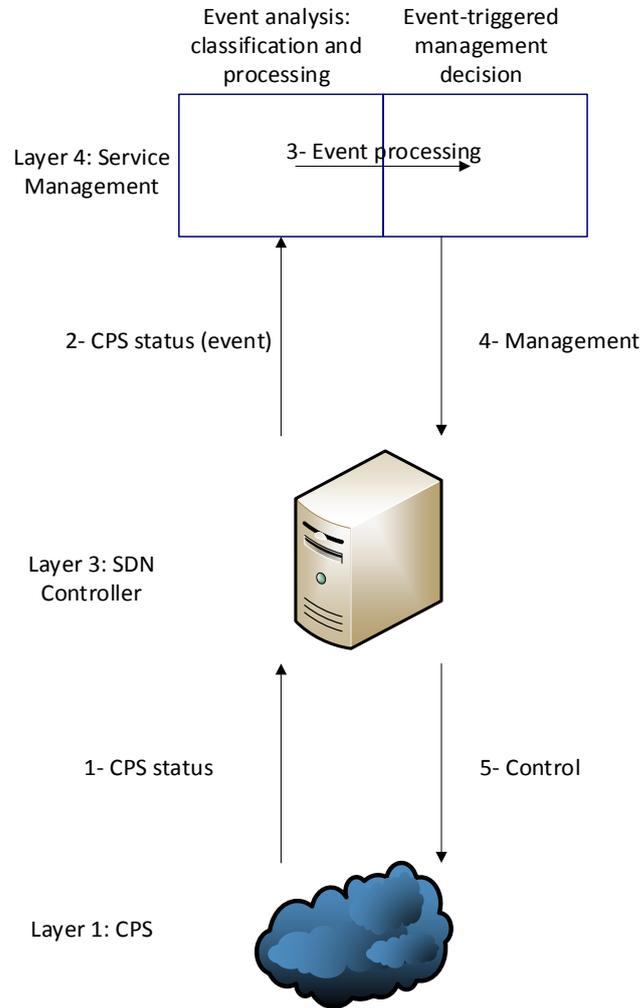

**Figure 2.** System Functional Blocks with monitor, classify, manage, and control messages

**Table II. Agent Event-Triggered Reactive Management Algorithm**

$S_0 = 1;\ \alpha = 0.8;\ n = 1$

**While True do**

    Collect, analyze, and classify CPS events occurred within last time slot

$$S = \frac{Quantity\_good\_events\_within\_last\_time\_slot}{Quantity\_total\_events\_within\_last\_time\_slot}$$

$S_n = S_{n-1} * \alpha + S * (1 - \alpha)$

    *if $S_n > threshold$ then*

        *CPS system is ok; do nothing different from last action*

    *else*

        *CPS system is not ok; play the cooperate / defect game*

    *end if*

    *n = n + 1*

*end for*

The next sub-section presents the basic design of a proactive software-defined resilient CPS management mechanism.

### 3.3. Design of a Proactive CPS Management Mechanism

This sub-section briefly debates how data analysis can also influence the CPS operation proactively. Figure 3 visualizes the CPS flowchart, showing the major functional phases of gathering data about the CPS operation (status), analyzing data, selecting a management decision, and applying the management decision on the CPS. After this iteration, more CPS data is gathered again, and the previous functional phases are repeated. We assume that data analysis can be performed using a machine learning algorithm to boost the system management [43]. In addition, the management decision of this data model should be conveniently matched with the event-triggered management decision of the agent discussed in sub-section 3.2. The orchestration among the two management methodologies (reactive vs proactive) can be made using a Blockchain solution [44], using a suitable consensus algorithm. Consensus algorithms, such as Kalman-based distributed algorithms [45], can provide interesting distributed functionalities of both filtering the menaces and manage CPSs to mitigate them (or even avoid them in the future). In this way, important network functions, e.g. firewall or Intrusion Detection/Prevention or honeypots, can be deployed pervasively within large networking edge domains, embedding a considerable number of sensors, actuators, or data aggregators. The authors of [45] discuss key recent results in the field of industrial CPSs modeled by differential dynamic equations, and they further discuss what issues should be addressed. These issues are analyzed from three distinct aspects: distributed filtering, distributed control and, distributed security control and filtering.

The next section models and analyzes a software-defined resilient CPS using an external mechanism to enforce cooperation among players through iterated repetitions of a non-cooperative game. The coordination among players can establish some orchestration among them towards a more efficient CPS global management, meaning the available CPS resources are efficiently used.

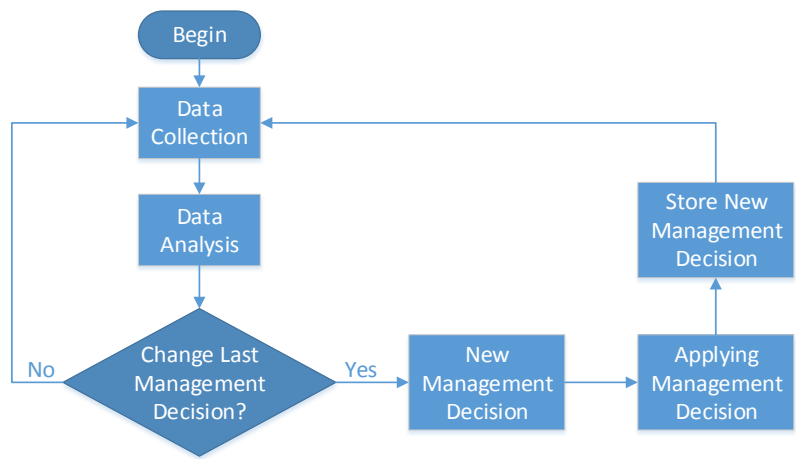

**Figure 3.** Data-triggered Management Model of a Cyber-Physical System

## 4. Analysis of a Software-Defined Resilient Cyber-Physical System

This section discusses a non-cooperative model enhanced by an external mechanism, for enforcing cooperation among players reactively through infinitely iterated repetitions of the model. The cooperation within players is very important for achieving reliable system operation with a limited set of resources. To support the next debate, we consider an infinitely repeated Prisoner's Dilemma (PD) game between two players. These players are two functional entities of a CPS and IoT system, e.g. containers, NFV, or specialized agents located at the topmost layer of the software-defined resilient CPS. By specialized agents we mean that, as the system overlaps a specific threshold, each system agent detects it and cooperatively reacts, selecting either 'absorb' or 'adapt to' the problem. Alternatively, the agent can be selfish by doing nothing to mitigate the problem.

Next, we analyze a model involving two topmost layer management agents that can either cooperate or defect. In the current game, the discounted factor combined with a mechanism that is triggered by a player´s defection can enforce cooperation throughout players. The discounted factor (<= 1) multiplies the payoffs of the current stage, meaning that in future game stages the payoffs of previous rounds have less relevance.

Figure 4 shows the payoff matrix of an infinitely repeated PD game as well as the total (per player accumulated) payoff along the initial four runs of the game, considering distinct values for the discounted factor (i.e. delta= {0,.2,.6,.95}). A Grim Trigger methodology is applied to a player that defects. Two distinct strategies are analyzed. In the first situation, both players cooperate, being rewarded with the social optimum payoff of 3 in every stage of the game, as shown in (1). In the second situation, one player defects in the first stage to increase its initial payoff from 3 to 5. Nevertheless, the other player at the second stage retaliates against the former defecting player, also defecting. Consequently, both players get a payoff of 1, as shown in (2), after the initial stage.

$$Coop = 3 + \delta.3 + \delta^2.3 + \cdots = \frac{3}{1-\delta} \tag{1}$$

$$Def = 5 + \delta.1 + \delta^2.1 + \cdots = 5 + \frac{\delta}{1-\delta} \tag{2}$$

$$\frac{3}{1-\delta} \geq 5 + \frac{\delta}{1-\delta} \Leftrightarrow \delta \geq 0.5 \tag{3}$$

The expression (3) evaluates the minimum value (i.e. 0.5) for the discounted factor (delta) to reflect in future a strong enough threat (in terms of payoff decrease) to a deviating player. Comparing the payoff trends of the two cases we have discussed in the previous paragraph, one can conclude for delta values of 0 (i.e. the game has only a single stage) and 0.2, which are both lower than 0.5, then the more convenient strategy for both players is always to defect (see Figure 4). The mutual defection occurs because the model gives the players solid evidence that they are playing the ultimate round of the game. So, the players are normally tempted to defect as they cannot be punished in the future. Alternatively, analyzing from Figure 4 the trends associated with the delta values of 0.6 and 0.95, which are both higher than 0.5, one can conclude that in the initial stages both players are tempted to defect; but after a threshold stage of the game is passed, both players should always cooperate in their best interest. This threshold stage depends on the delta value (see Figure 4). In fact, as the delta value increases towards one that means the player (with that perspective of the game) learns it is better to cooperate instead defecting faster, i.e. after fewer stages counted from the game's start.

The opposite happens if for the same game the strategy is changed from Tit for Tat (Figure 4) to Slow Tit for Two Tats (Figure 5). From Figure 5 it is evident that the need to cooperate occurs in later stages of the game when compared with the trend of Figure 4. The last difference in behavior occurs because Tit for Two Tats is a forgiving strategy by which a player only defects after the opponent has defected twice in a row. This behavior is fairer than Tit for Tat in scenarios where the player, due to a network communication error or any limitation imposed by other system operational constrain, erroneously perceived the previous opponent's choice.

For validating the key conclusions extracted from the previous analytical comparison made between the two cooperative strategies, we have performed some additional simulations to study how the distribution of the two studied types of cooperative behavior evolve along the time, considering a total player population of constant size. In this way, Figure 6 shows the evolution of a population composed by the two types of players of our study, during one thousand rounds. The Moran process was used to keep the population size always at a constant value of one hundred players. For simulating that situation, we have used the Axelrod Python library[1]. The winning strategy was Slow Tit for Two Tats, suggesting Tit for Tat has a lower fitness function than the former one, confirming our analytical conclusions.

There are still a considerable number of open research issues in discounted repeated games, such as follows: i) study scenarios with myopic information; ii) payoff variations (controlling players defection); iii) considering a distinct delta per player; iv) triggering punishment only during a limited set of stages (i.e. forgiving after T stages following a deviating behavior); v) studying scenarios where the end of the game at the current round (with probability 1-delta) due to the absence of cooperation among players could alternatively be related to the probability of the occurrence of a serious system threat at the current round, stopping the system's operation and the game. Also, the authors of [46] analyze a metadata set of experiments on infinitely repeated PD games, reaching some promising results. Their results suggest that cooperation is affected by infinite repetition and is more likely to arise when it occurs in equilibrium. However, the fact that cooperation can be supported in equilibrium does not imply that most players will cooperate. High cooperation rates will emerge only when the parameters of the repeated game are such that cooperation is very robust to strategic uncertainty. Another recent contribution [47] suggests the usage of statistical physics to understand human cooperation better. We think that this research direction is very interesting to transpose to an investigation on how to design and deploy systems used by diverse players, which need to be more cooperative and fairer in the mutual interaction within each system. Further, the cooperation and fairness should be obtained not by centralized policies that can often be either unoptimized or be easily deceived by (some) players, but simply by a higher level of collective intelligence stemming from each player.

The current section has used static rewards for the strategies each player is able to select. It is also possible to analyze a more dynamic game, where the reward of each player is a function of the estimator $S_n$ used by the algorithm of Table II.

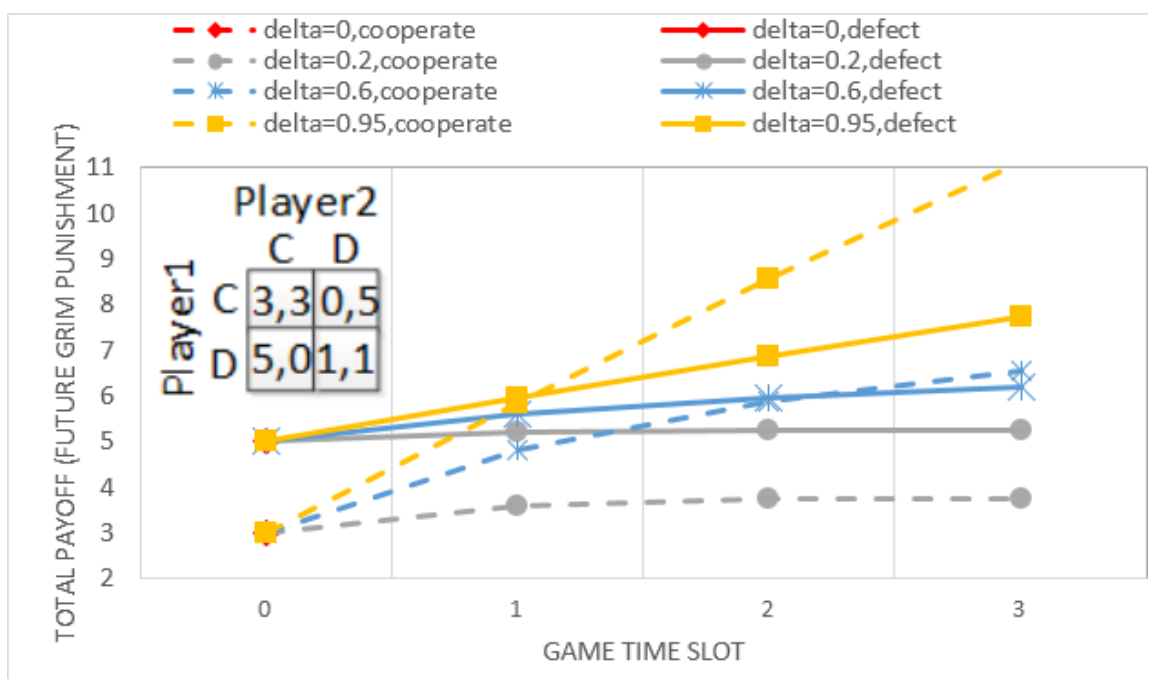

**Figure 4.** Total Payoff Trend for a specific player involved in an Infinitely Repeated PD Game with a Grim Punishment Mechanism (Tit for Tat) and Diverse Discounted Factors

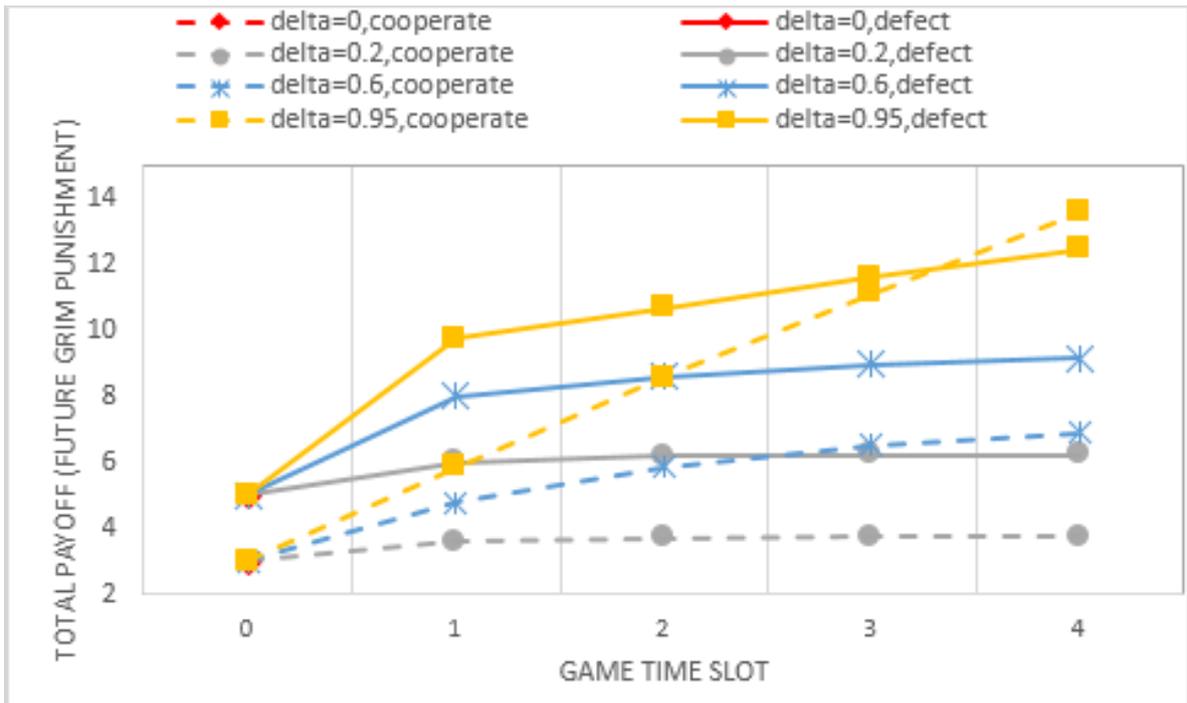

**Figure 5.** Total Payoff Trend for a specific player involved in an Infinitely Repeated PD Game with a Grim Punishment Mechanism (Slow Tit for Two Tats) and Diverse Discounted Factors

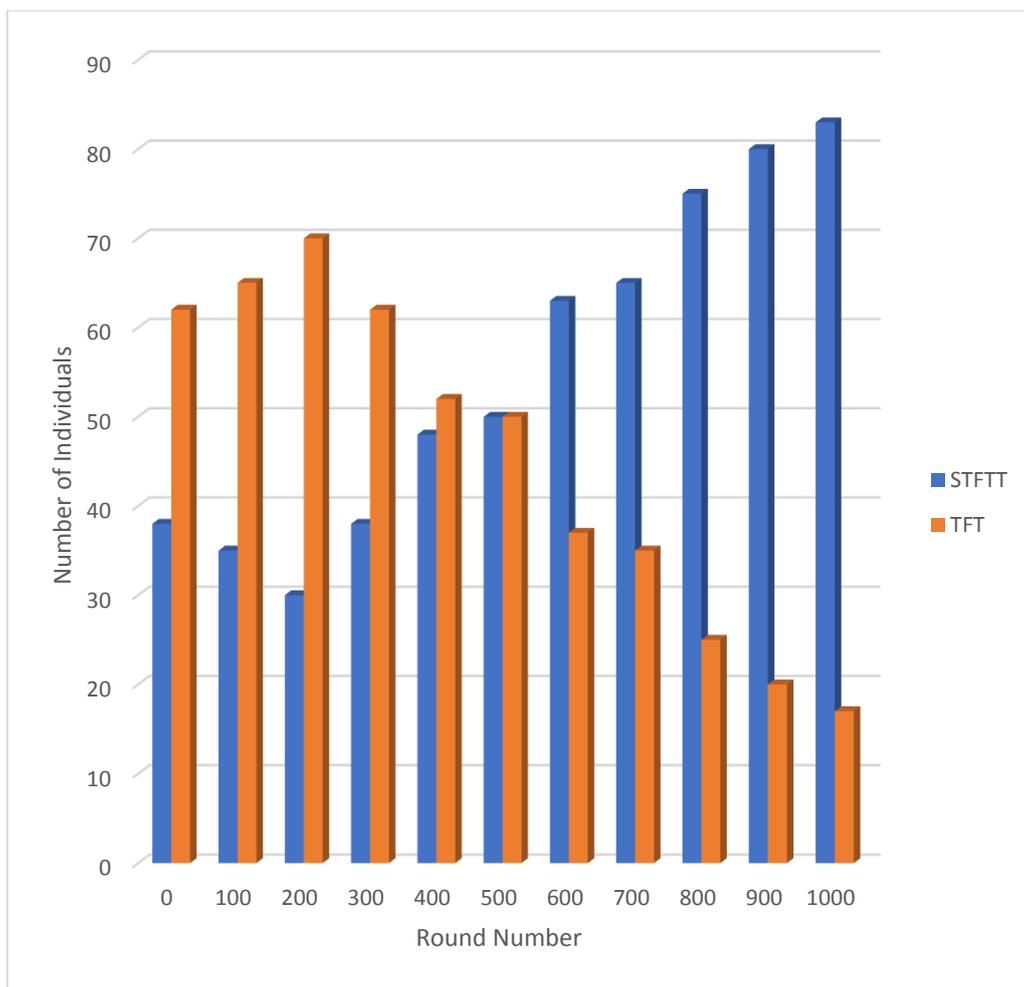

**Figure 6.** Distribution of the Two Studied Types of Cooperative Behavior in a Constant Population of 100 Players during 1,000 Rounds (STFTT – Slow Tit For Two Tats; TFT – Tit For Tat)

## 5. Conclusion and Future Work

Cyber-Physical Systems (CPSs) are increasingly important in very important areas of our society but are subject to new challenges to their optimum and reliable operation, which strongly suggests the need for innovative management strategies to achieve resilience in these systems. To this end, we have proposed an SDN-based solution based on distributed Network Function Virtualization (NFV) modules. These NFV modules take orchestrated management decisions among themselves to ensure a resilient CPS configuration against threats and an optimum operation of the CPSs. In this paper, we design, model and compare distinct incentive mechanisms to enforce cooperation among NFVs.

Future work will involve studying data-triggered management models for building programmable resilient and smart CPSs [48]. In addition, the dynamic movement of processes and data in (edge) cloud-based systems may compromise their resilience, unless steps are taken to recognize the problem and modify anomaly detection components appropriately [49].

**Funding:** The work of Jose Moura is funded by FCT/MCTES through national funds and when applicable co-funded EU funds under the project UIDB/EEA/50008/2020.

**Acknowledgments:** Jose Moura acknowledges the support given by Instituto de Telecomunicações, Lisbon, Portugal. David Hutchison is grateful to colleagues in EU COST Action RECODIS (CA15127) for discussions about the resilience of communication systems.